\documentclass[12pt]{article}
\usepackage{graphicx}
\DeclareGraphicsExtensions{.pdf}
\usepackage{float} 
\usepackage{amsmath}
\usepackage{amsfonts}   
\usepackage{amssymb}

\begin{document}
\date{}
\title{{\bf{\Large Holographic thermalization from non relativistic branes}}}
\author{
 {\bf {\normalsize Dibakar Roychowdhury}$
$\thanks{E-mail:  dibakarphys@gmail.com, dibakarr@iitk.ac.in}}\\
 {\normalsize  Indian Institute of Technology, Department of Physics,}\\
  {\normalsize Kanpur 208016, Uttar Pradesh, India}
}

\maketitle
\begin{abstract}
In this paper, based on the fundamental principles of Gauge/gravity duality and considering a \textit{global quench}, we probe the physics of thermalization for certain special classes of strongly coupled non relativistic QFTs those are dual to an asymptotically Schr{\"o}dinger $ Dp $ brane space time. In our analysis, we note that during the pre local stages of the thermal equilibrium the entanglement entropy has a faster growth in time compared to its relativistic cousin. However, it shows a linear growth during the post local stages of thermal equilibrium where the so called tsunami velocity associated with the linear growth of the entanglement entropy saturates to that of its value corresponding to the relativistic scenario. Finally, we explore the saturation region and it turns out that one must constraint certain parameters of the theory in a specific way in order to have discontinuous transitions at the point of saturation.
\end{abstract}

\section{Overview and Motivation}
One of the most recent as well as promising developments that took place during the last couple of years is the understanding of non equilibrium dynamics in strongly interacting Quantum Field Theories (QFTs) at finite temperatures. For quantum mechanical systems in equilibrium, it is the RG flow that helps us to identify various universal features those are insensitive to the microscopic details of the system. For systems out of equilibrium, such notion of universality does not hold in general and in fact one does not have much options to deal with when the system itself is strongly coupled. Under such circumstances, the so called AdS/CFT duality \cite{Maldacena:1997re}-\cite{Witten:1998qj} turns out to be an extremely elegant tool in order to explore various universal features among strongly interacting QFTs.

In order to study non equilibrium dynamics in strongly interacting QFTs, one typically starts with a system that is primarily in some global equilibrium configuration that could be either its ground state at $ T=0 $ or some thermally excited state at finite temperatures. The natural next step would be to put this QFT out of its global equilibrium by turning on sources for some (relevant) operator in the theory namely\footnote{In this paper we would be considering the case where, $ \varrho(x) = \varrho(\mathfrak{t}) $ which is thereby termed as \textit{global quench}.},
\begin{eqnarray}
S_{QFT} \rightarrow S_{QFT} + \int d^{D}x \varrho (x) \mathcal{O} (x).
\end{eqnarray}

During the process of thermalization, the so called thermodynamic variables such as the pressure, temperature etc. are no more valid entities and one might try to probe such systems in terms of various \textit{non local} observable like the entanglement entropy of the system. However, for strongly coupled systems such observable are hard to compute directly from the field theory perspective and therefore one needs to rely on various holographic techniques \cite{Liu:2013qca}-\cite{Balasubramanian:2010ce}.

The process that we describe in this paper, essentially corresponds to a rapid thermalization followed by a global quench where one injects a uniform density of matter for a very short interval of time ($ \delta \mathfrak{t}\sim 0 $). Holographically, such processes are described in terms of the process of black hole formation during the gravitational collapse of a thin shell of matter in an AdS-Vaidya space time \cite{Liu:2013qca},
\begin{eqnarray}
ds^{2}=\frac{L^{2}}{u^{2}}(-f(u,\upsilon)d\upsilon^{2}-2 d\upsilon du +d\textbf{x}^{2})
\end{eqnarray}
where, $ \upsilon \sim \mathfrak{t}-u $ is the so called in going null coordinate that typically plays the role of time for the theory living on the boundary ($ u \sim 0 $). In order to describe rapid thermalization, the function $f(u,\upsilon)$ must take the following form,
\begin{eqnarray}
f(u,\upsilon) =1 -\Theta (\upsilon)h(u)
\label{ee1}
\end{eqnarray}
where, $ \Theta (\upsilon) $ is the so called step function, such that for $ \upsilon <0 $, in the dual gravity picture we have a pure AdS space time which corresponds to a ground state of the dual field theory, whereas on the other hand, for $ \upsilon >0 $, the corresponding dual gravity picture encodes a black hole space time which describes the dual QFT in its thermally excited state \cite{Liu:2013qca}. 

As far as the motivation is concerned, the purpose of the present analysis is to understand the physics of thermalization in a non relativistic set up where the underlying symmetry group of the dual field theory is characterized by the Schr{\"o}dinger (Sch) algebra rather than the usual relativistic conformal algebra. In the recent years, QFTs with Schr{\"o}dinger isometry group has gained renewed attention in the context of the celebrated AdS/cold atom correspondence \cite{Son:2008ye}-\cite{Balasubramanian:2008dm} which states that the Schr{\"o}dinger isometry group corresponding to a system of fermions at unitarity could possibly be realized as an isometry group of some dual gravitational theory in the bulk \cite{Adams:2008wt}-\cite{Roychowdhury:2014lta}. 
The question that naturally arises in this context is that what is the fate of thermalization for such a system of fermions at unitarity and this issue has never been addressed in the literature to the best of our knowledge. The goal of the present article is to fill up this gap and provide some insights into the physics of thermalization corresponding to a system of fermions at unitarity.

 To be more precise, suppose we start with some non relativistic quatum mechanical system (that preserves Schr{\"o}dinger algebra) in its ground state which suddenly excites and subsequently thermalizes following a \textit{global} quench. What we would like to understand is how one could possibly probe such non equilibrium processes by computing the entanglement entropy of the system. In other words, we would like to understand how does the entanglement entropy corresponding to such systems evolve in time and consequently saturate at late times \cite{Calabrese:2005in}-\cite{Calabrese:2009qy}. In order to model such quench processes in a holographic setup, in the dual gravitational counterpart one needs to consider some non relativistic metric of the above form (\ref{ee1}) where one should expect an instantaneous transition from the pure Schr{\"o}dinger space time to the Schr{\"o}dinger black brane space time for $ \upsilon \geq 0$.  

The organization of the rest of the paper is the following: In Section 2, we provide details regarding our holographic construction in the bulk and compute the entanglement entropy following the holographic prescription \cite{Ryu:2006bv}-\cite{Ryu:2006ef}. Finally, we conclude in Section 3.

\section{Thermalization}
\subsection{Basic set up}
We start our analysis with a formal introduction to the non relativistic set up in the bulk which is essentially described in terms of a five dimensional \textit{uncharged} Schr{\"o}dinger black brane ($Sch_5 $) solution of the following form \cite{Herzog:2008wg}, \cite{Roychowdhury:2014lta},
\begin{eqnarray}
ds_{5}^{2}&=&\frac{\mathcal{K}^{1/3}L^{2}}{u^{2}}\left[ - \frac{f}{\mathcal{K}}d\tau^{2}-\frac{f \beta^{2}L^{4}}{u^{2}\mathcal{K}}(d\tau + d\chi)^{2}+ \frac{d\chi^{2}}{\mathcal{K}}  + \frac{du^{2}}{f}+dx^{2}+dy^{2}  \right] \nonumber\\
f(u)&=&1-\left(\frac{u}{u_{H}} \right)^{4},~~\mathcal{K}(u)= 1+\frac{\beta^{2}L^{4}u^{2}}{u_H^{4}} \label{E1}  
\end{eqnarray}
where, the horizon of the black brane is located at $ u=u_H $, whereas, on the other hand the boundary of the spacetime is located at $ u=0 $. Here, $ \beta $ is a dimensionful quantity (with dimension, $ [\beta] = \frac{1}{L}$) associated with the Null Melvin Twist and is related to the particle number in the dual field theory. The coordinates, $ \tau $ and $ \chi $ are related to the isometry directions of the original type IIB supergravity solution in 10 dimensions \cite{Herzog:2008wg}. 

In order to proceed further, our next step would be to search for the corresponding Eddington-Finkelstein like representation of the above black brane configuration (\ref{E1}). In order to do so, we introduce the light cone time $ \upsilon $ as,
\begin{eqnarray}
d \upsilon = d\tau - \sqrt{\frac{\mathcal{K}}{1+\frac{\beta^{2}L^{4}}{u^{2}}}}\frac{du}{f}.
\label{E2}
\end{eqnarray}

Substituting (\ref{E2}) into (\ref{E1}) we obtain,
\begin{eqnarray}
ds_{5}^{2}=\frac{\mathcal{K}^{1/3}L^{2}}{u^{2}}\left[ - \frac{\tilde{f}}{\mathcal{K}} d\upsilon^{2}-2 h(u) d\upsilon du-\frac{2f\beta^{2}L^{4}}{u^{2}\mathcal{K}}d\upsilon d\chi - \frac{2 \beta^{2}L^{4}}{u^{2}\mathcal{K}h}du d\chi + c(u)d\chi^{2}+ dx^{2}+dy^{2}  \right]
\label{E3}
\end{eqnarray}
where,
\begin{eqnarray}
\tilde{f}(u)&=&f(u)\left(1+ \frac{\beta^{2}L^{4}}{u^{2}}\right)\equiv  f(u)g(u);~~g(u)= \left(1+ \frac{\beta^{2}L^{4}}{u^{2}}\right) \nonumber\\
h(u)&=&\sqrt{\frac{g(u)}{\mathcal{K}(u)}}\nonumber\\
c(u)&=&\frac{1}{\mathcal{K}(u)}\left(1-\frac{f(u)\beta^{2}L^{4}}{u^{2}} \right).  
\end{eqnarray}

In the limit of the vanishing time interval ($ \Delta t \sim 0 $) associated with external sources at the boundary, the width of the collapsing shell in the bulk eventually goes to zero \cite{Liu:2013qca}. Under such circumstances, the functions defined above (\ref{E3}) could be replaced as,
\begin{eqnarray}
f(u , \upsilon)&=&1- \Theta (\upsilon)\left(\frac{u}{u_{H}} \right)^{4}\nonumber\\
\mathcal{K}(u, \upsilon)&=&  1+ \Theta (\upsilon)\left(  \frac{\beta^{2}L^{4}u^{2}}{u_H^{4}}\right) \nonumber\\
\tilde{f}(u, \upsilon)&=&   f(u , \upsilon) g(u)\nonumber\\
h(u, \upsilon)&=&\sqrt{\frac{g(u)}{\mathcal{K}(u, \upsilon)}}\nonumber\\
c(u, \upsilon)&=& \frac{1}{\mathcal{K}(u, \upsilon)} \left(1-\frac{f(u, \upsilon)\beta^{2}L^{4}}{u^{2}} \right). 
\end{eqnarray}
where, $ \Theta (\upsilon) $ is the so called step function, such that for, $ \upsilon <0 $ we are left with a pure Schr{\"o}dinger space time of the form, 
\begin{eqnarray}
ds_{5}^{2}=\frac{L^{2}}{u^{2}}\left[ -g(u)  d\upsilon^{2}-2\sqrt{g} d\upsilon du-\frac{2\beta^{2}L^{4}}{u^{2}}d\upsilon d\chi - \frac{2 \beta^{2}L^{4}}{u^{2}\sqrt{g}}du d\chi + \left(1- \frac{\beta^{2}L^{4}}{u^{2}}\right)d\chi^{2}+ d\textbf{x}^{2} \right]
\end{eqnarray}
whereas, on the other hand, for $ \upsilon > 0 $ the space time turns out to be precisely that of the $Sch_5 $ black brane configuration (\ref{E3}).

\subsection{Area functional: Preliminaries}
The purpose of the present section is to explore the so called holographic entanglement entropy (HEE) \cite{Ryu:2006bv}-\cite{Ryu:2006ef} as a candidate that probes the dynamics of thermal quench in a strongly coupled non relativistic plasma living on the conformal boundary of $Sch_5 $ space time. In order to compute the HEE, we  consider a (rectangular) strip times $ \chi $ like region at the boundary namely \cite{Ryu:2006bv}-\cite{Ryu:2006ef},
\begin{eqnarray}
-\mathfrak{R}/2 \leq x \leq \mathfrak{R}/2;~~~0 \leq y \leq L_{y};~~0\leq \chi \leq L_{\chi}
\end{eqnarray}
and compute the area functional associated with the extremal surface that extends into the bulk. Note that, here $ L_{\chi} (=\int d\chi)$, is the length scale associated with the $ \chi $ direction\cite{Kim:2012nb}. 

Since in the present example, we are dealing with a time dependent scenario, therefore it would be fare enough to use the so called covariant proposal for HEE \cite{Hubeny:2007xt}. This in turn suggests that one should parametrize the so called extremal surface in terms of the functions, $ \upsilon (x) $ and $ u(x) $. The boundary conditions that one should impose under such circumstances are given by the following set of constraints,
\begin{eqnarray}
u(\mathfrak{R}/2)=0,~~\upsilon (\mathfrak{R}/2)= \mathfrak{t},~~u'(0)=\upsilon'(0)=0\nonumber\\
u(0)=u_T,~~\upsilon(0)=\upsilon_{T}
\end{eqnarray} 
where, $ u_T $ and $ \upsilon_{T} $ correspond to the values of the functions at the tip of the extremal surface in the bulk.
\subsubsection{Equations of motion}
In order to evaluate the equations of for the extremal surface in the bulk, we first note down the induced metric on the extremal surface,
\begin{eqnarray}
ds^{2}_{ES}=\frac{\mathcal{K}^{1/3}L^{2}}{u^{2}}\left[ \left( 1 - \frac{\tilde{f}(u,\upsilon)}{\mathcal{K}(u, \upsilon)} \upsilon'^{2}-2 h(u,\upsilon) \upsilon' u'\right) dx^{2}-\frac{2\beta^{2}L^{4}}{u^{2}\mathcal{K}}\left(f \upsilon' + \frac{u'}{h}\right) dx d\chi + c(u, \upsilon)d\chi^{2}+ dy^{2}  \right].
\label{E9}
\end{eqnarray}

Using (\ref{E9}), the corresponding area functional finally turns out to be,
\begin{eqnarray}
\mathcal{A}=L_{y}L_{\chi}L^{3}\int_{-\mathfrak{R}/2}^{\mathfrak{R}/2}dx \frac{\sqrt{\mathcal{K}(u,\upsilon)\mathsf{Q}(u,\upsilon)}}{u^{3}}
\label{E10}
\end{eqnarray}
where the function $ \mathsf{Q}(u,\upsilon) $ could be formally expressed as,
\begin{eqnarray}
\mathsf{Q}(u,\upsilon) = c(u, \upsilon)-\frac{\upsilon'^{2}f(u,\upsilon)}{\mathcal{K}(u,\upsilon)}\left(g(u)c (u,\upsilon) +\frac{\beta^{4}L^{8}f(u,\upsilon)}{u^{4}\mathcal{K}(u, \upsilon)}\right)-\frac{\beta^{4}L^{8}u'^{2}}{u^{4}h^{2}(u,\upsilon)\mathcal{K}^{2}(u,\upsilon)}\nonumber\\
- 2 \upsilon' u' \left(  h (u, \upsilon)c(u, \upsilon)+\frac{\beta^{4}L^{8}f(u,\upsilon)}{u^{4}\mathcal{K}^{2}(u, \upsilon)h(u,\upsilon)}\right). 
\label{E11} 
\end{eqnarray}

The equations of motion, that readily follow from (\ref{E10}), could be formally expressed as,
\begin{eqnarray}
u^{3}\sqrt{\mathcal{G}(u,\upsilon)}\partial_{x}\left( \frac{\upsilon' f\left( g(u)c(u,\upsilon)+\frac{f\beta^{4}L^{8}}{u^{4}\mathcal{K}}\right)+\mathcal{K}u'\left(  h (u, \upsilon)c(u, \upsilon)+\frac{\beta^{4}L^{8}f}{u^{4}\mathcal{K}^{2}h}\right) }{u^{3}\sqrt{\mathcal{G}(u,\upsilon)}}\right) &=&\frac{1}{2}\mathcal{F}_{\upsilon}(u,\upsilon)\nonumber\\
u^{3}\sqrt{\mathcal{G}(u,\upsilon)}\partial_{x}\left( \frac{\frac{\beta^{4}L^{8}u'}{u^{4}h^{2}\mathcal{K}(u,\upsilon)}+\mathcal{K}\upsilon'\left(  h (u, \upsilon)c(u, \upsilon)+\frac{\beta^{4}L^{8}f}{u^{4}\mathcal{K}^{2}h}\right)}{u^{3}\sqrt{\mathcal{G}(u,\upsilon)}}\right)-\frac{3 \mathcal{G}(u, \upsilon)}{u}& =&\frac{1}{2}\mathcal{F}_{u}(u,\upsilon)\nonumber\\
\label{E12}
\end{eqnarray} 
where,
\begin{eqnarray*}
\mathcal{G}(u, \upsilon)=\mathcal{K}(u,\upsilon)\mathsf{Q}(u,\upsilon),
\end{eqnarray*}
\begin{eqnarray*}
\mathcal{F}_{\upsilon}(u,\upsilon) =\frac{\partial f}{\partial \upsilon} \upsilon'^{2}g(u)c(u,\upsilon)-\frac{\partial \mathcal{K}}{\partial \upsilon}\mathsf{Q}(u,\upsilon)-\frac{\partial c}{\partial \upsilon}\mathcal{K}(u, \upsilon)-\frac{\partial \mathcal{K}}{\partial \upsilon}\frac{\upsilon'^{2}f(u, \upsilon)}{\mathcal{K}(u,\upsilon)}g(u)c(u,\upsilon)\nonumber\\
+\frac{\partial c}{\partial \upsilon}\upsilon'^{2}g(u)f(u,\upsilon)+\frac{2f(u,\upsilon) \upsilon'^{2}\beta^{4}L^{8}}{u^{4}\mathcal{K}(u, \upsilon)}\left( \frac{\partial f(u,\upsilon)}{\partial \upsilon}-\frac{f(u,\upsilon)}{\mathcal{K}(u,\upsilon)}\frac{\partial \mathcal{K}(u,\upsilon)}{\partial \upsilon}\right)\nonumber\\
-\frac{2u'^{2}\beta^{4}L^{8}}{u^{4}h^{2}(u,\upsilon)\mathcal{K}(u,\upsilon)}\left( \frac{1}{h}\frac{\partial h}{\partial \upsilon}+\frac{1}{\mathcal{K}}\frac{\partial \mathcal{K}}{\partial \upsilon}\right)+2\mathcal{K}(u,\upsilon)\upsilon' u' \frac{\partial}{\partial \upsilon}(h(u,\upsilon)c(u,\upsilon))\nonumber\\
+\frac{2 \upsilon' u' \beta^{4}L^{8}}{u^{4}\mathcal{K}(u,\upsilon)h(u,\upsilon)}\left( \frac{\partial f}{\partial \upsilon}-\frac{2 f(u,\upsilon)}{\mathcal{K}(u, \upsilon)}\frac{\partial \mathcal{K}}{\partial \upsilon}-\frac{ f(u,\upsilon)}{h(u, \upsilon)}\frac{\partial h}{\partial \upsilon}\right),  
\end{eqnarray*}
\begin{eqnarray*}
\mathcal{F}_{u}(u,\upsilon) =\frac{\partial f}{\partial u} \upsilon'^{2}g(u)c(u,\upsilon)-\frac{\partial \mathcal{K}}{\partial u}\mathsf{Q}(u,\upsilon)-\frac{\partial c}{\partial u}\mathcal{K}(u, \upsilon)+\frac{\partial (c(u,\upsilon)g(u))}{\partial u}\upsilon'^{2}f(u,\upsilon)\nonumber\\
-\frac{\partial \mathcal{K}}{\partial u}\frac{\upsilon'^{2}f(u, \upsilon)}{\mathcal{K}(u,\upsilon)}g(u)c(u,\upsilon)+\frac{2f(u,\upsilon) \upsilon'^{2}\beta^{4}L^{8}}{u^{4}\mathcal{K}(u, \upsilon)}\left( \frac{\partial f(u,\upsilon)}{\partial u}-\frac{f(u,\upsilon)}{\mathcal{K}(u,\upsilon)}\frac{\partial \mathcal{K}(u,\upsilon)}{\partial u}\right)\nonumber\\
-\frac{4\upsilon'^{2}f^{2}(u,\upsilon) \beta^{4}L^{8}}{\mathcal{K}(u,\upsilon)u^{5}}-\frac{2u'^{2}\beta^{4}L^{8}}{u^{4}h^{2}(u,\upsilon)\mathcal{K}(u,\upsilon)}\left( \frac{1}{h}\frac{\partial h}{\partial u}+\frac{1}{\mathcal{K}}\frac{\partial \mathcal{K}}{\partial u}+\frac{2}{u}\right)+2\mathcal{K}\upsilon' u' \frac{\partial}{\partial u}(h(u,\upsilon)c(u,\upsilon))\nonumber\\ 
+\frac{2 \upsilon' u' \beta^{4}L^{8}}{u^{4}\mathcal{K}(u,\upsilon)h(u,\upsilon)}\left( \frac{\partial f}{\partial u}-\frac{2 f(u,\upsilon)}{\mathcal{K}(u, \upsilon)}\frac{\partial \mathcal{K}}{\partial u}-\frac{ f(u,\upsilon)}{h(u, \upsilon)}\frac{\partial h}{\partial u}-\frac{4 f(u, \upsilon)}{u}\right).  
\end{eqnarray*}

Our next step would be compute the first integral of motion. Since the integrand in (\ref{E10}) does not explicitly depend on $ x $, therefore the first integral of motion turns out to be,
\begin{eqnarray}
\frac{u^{3}\sqrt{\mathcal{G}(u,\upsilon)}}{\mathcal{K}(u,\upsilon) c(u,\upsilon)}=\mathsf{J}=const.
\label{E13}
\end{eqnarray}

Before we proceed further, it is customary to note that the extremal surface that we consider in our analysis extends both in the pure $Sch_5 $ as well as in the $Sch_5 $ black brane regions. Moreover, by considering the reflection symmetry of the entire configuration around $ x=0 $, it is indeed sufficient to consider only the positive half ($ x>0 $) of the entangling region at the boundary. In the following, we discuss both $ \upsilon <0 $, as well as, $ \upsilon >0 $ regions separately.
\subsubsection{Region I: $ \upsilon < 0 $}
In the region $ \upsilon < 0 $, the first one of the above set of equations (\ref{E12}) enormously simplifies to give,
\begin{eqnarray}
\frac{1}{c(u)}\left(\upsilon' +\frac{u'}{\sqrt{g(u)}} \right)=\mathsf{E}.
\label{E14}
\end{eqnarray} 

Under such circumstances, using (\ref{E14}), one could further simplify (\ref{E11}) as,
\begin{eqnarray}
\mathsf{Q}(u)= c(u)\left( 1+u'^{2} \right).
\label{E15}
\end{eqnarray} 

Using (\ref{E15}), it is quite trivial to show that \footnote{Note that, here we have considered the fact that, $ u'(x)<0 $ throughout its entire domain of definition.},
\begin{eqnarray}
u'= -\frac{\sqrt{c(u)}}{u^{3}}\sqrt{\mathsf{J}^{2}-\frac{u^{6}}{c(u)}}
\end{eqnarray}
which could be further integrated to obtain,
\begin{eqnarray}
x(u)=\int_{u}^{u_T}\frac{u'^{3}du'}{\sqrt{c(u')}}\frac{1}{\sqrt{\mathsf{J}^{2}-\frac{u'^{6}}{c(u')}}}.
\label{E17}
\end{eqnarray}

Finally, from (\ref{E14}), it is also trivial to show that,
\begin{eqnarray}
\upsilon = \upsilon_{T}+u_T \sqrt{1+\frac{\beta^2 L^4}{u_T^2}} -u \sqrt{1+\frac{\beta^2 L^4}{u^2}}.
\label{E18}
\end{eqnarray}

\subsubsection{Matching on the shell}
Before we proceed further, let us define the values corresponding to $ u $ and $ x $ at the point of intersection with the null shell, $ \upsilon =0 $ as $ u_c $ and $ x_c $ respectively which finally yields,
\begin{eqnarray}
u_c \sqrt{1+\frac{\beta^2 L^4}{u_c^2}} = u_T \sqrt{1+\frac{\beta^2 L^4}{u_T^2}}+\upsilon_{T}.
\end{eqnarray}

On the other hand, taking derivatives on both sides of (\ref{E18}) we find\footnote{Here, $ - $ and $ + $ subscripts refer to entities in the regions, $ \upsilon <0 $ and $ \upsilon >0 $ respectively.},
\begin{eqnarray}
u'_{-}=-\sqrt{g(u_c)}~\upsilon'_{-}=-\frac{\sqrt{c(u_{c})}}{u_{c}^{3}}\sqrt{\mathsf{J}^{2}-\frac{u_{c}^{6}}{c(u_c)}}.
\label{E20}
\end{eqnarray}

To find the derivatives on the other side of the null shell, our next task would be to integrate (\ref{E12}) across the null shell which finally enable us to establish the precise map between different kinematics on the both sides of the null shell. 

In order to proceed further, let us first note that since we inject matter along the null direction $ \upsilon $, therefore the corresponding conjugate momentum must encounter a jump as we move from the region, $ \upsilon <0 $ to the region, $ \upsilon>0 $. On the other hand, the momentum conjugate to $ u $ must remain continuous across the null shell, $ \upsilon =0 $ \cite{Alishahiha:2014cwa}. The second condition naturally implies that,
\begin{eqnarray}
\mathsf{P}_{-}^{(u)}=\mathsf{P}_{0}^{(u)}=\mathsf{P}_{+}^{(u)}
\end{eqnarray}
where, $ \mathsf{P}_{0}^{(u)} $ stands for the momentum (conjugate to $ u $) exactly on the null shell, $ \upsilon =0 $. Now, computing momenta corresponding to the regions, $ \upsilon <0 $ and $ \upsilon>0 $ and taking the limit, $ \upsilon \rightarrow 0 $ we finally arrive,
\begin{eqnarray}
 \upsilon'_{-}&=&\upsilon'_{+}+\frac{\beta^{4}L^{8}}{u_c^{4}\sqrt{g(u_c)}}(u'_{+}-u'_{-}).
 \label{E22}
\end{eqnarray}

As a next step of our analysis, we first focus on the first equation in (\ref{E12}). After performing the integration in the vicinity of the null shell, $ \upsilon =0 $ we find,
\begin{eqnarray}
u'_{+}=u'_{-}-\frac{\mathcal{Z}(u_c)}{2\sqrt{g(u_c)}c(u_c)}
\label{E23}
\end{eqnarray}
where, the function $ \mathcal{Z}(u_c) $ could be formally expressed as,
\begin{eqnarray}
\mathcal{Z}(u_c) =u'_{-}\sqrt{g(u_c)}c(u_c)\left( \frac{u_c}{u_{H}}\right)^{4} -\frac{\beta^{2}L^{4}u_c^{2}}{\upsilon'^{(0)}u_H^{4}}
-\frac{\sqrt{g(u_c)}|u'_c| u_c^{2}\beta^{2}L^{4}}{u_H^{4}}.
\label{E24}
\end{eqnarray}
Note that, here $ \upsilon'^{(0)} $ and $ u'_c $ correspond to the derivatives exactly on the null shell, $ \upsilon =0 $. Substituting (\ref{E24}) into (\ref{E23}), one could further simplify the resulting expression as,
\begin{eqnarray}
u'_{+}=\left( 1-\frac{1}{2}\left( \frac{u_c}{u_{H}}\right)^{4}\right)u'_{-} + \frac{\beta^{2}L^{4}u_c^{2}}{2\sqrt{g(u_c)}c(u_c)u_H^{4}}\left( \frac{1}{\upsilon'^{(0)}}+\sqrt{g(u_c)}|u'_c|\right).
\label{e28}
\end{eqnarray}

Finally, after some trivial algebra it is indeed quite straightforward to show,
\begin{eqnarray}
\mathcal{G}_{+}-\mathcal{G}_{-}=\mathsf{Q}_{+}-\mathsf{Q}_{-}=c(u_c)(u'^{2}_{+}-u'^{2}_{-}).
\end{eqnarray}

\subsubsection{Region II: $ \upsilon > 0 $}
We now turn our attention towards the black brane sector of the entire space time configuration. As a first step of our analysis, using (\ref{E20}), (\ref{E22}) and (\ref{E24}) and considering the close vicinity of the null shell we find,
\begin{eqnarray}
\mathsf{E}=-\frac{\sqrt{g(u_c)}}{2}|u'_{-}|\left( \frac{u_c}{u_{H}}\right)^{4} +\frac{\beta^{2}L^{4}u_c^{2}}{2c(u_c)u_H^{4}}\left( \frac{1}{\upsilon'^{(0)}}+\sqrt{g(u_c)}|u'_c|\right).
\label{E27}
\end{eqnarray}
Therefore, unlike the previous examples in the literature \cite{Liu:2013qca}, the first integral of motion is not guaranteed to be negative for non relativistic background.

We now focus on the first equation in (\ref{E12}), which yields the following relation,
\begin{eqnarray}
\frac{1}{\mathcal{K}(u)c(u)}\left(\frac{u'}{h(u)}+\upsilon' f(u) \right) = \mathsf{E}
\end{eqnarray}
which could be further simplified in order to obtain,
\begin{eqnarray}
\upsilon' = \frac{1}{f(u)}\left( \mathsf{E}-\frac{u'}{h(u)}\right)-\frac{\beta^{2}L^{4}\mathsf{E}}{u^{2}}.
\label{E29} 
\end{eqnarray}

Using (\ref{E13}) and (\ref{E29}), it is indeed quite straightforward to find,
\begin{eqnarray}
u'^{2}=f(u)\left( \frac{\mathsf{J}^{2}}{u^{6}}-1\right) + \mathsf{E}^{2}\left(1-\frac{f(u)\beta^{2}L^{4}}{u^{2}} \right) \equiv \mathcal{H}(u)
\label{E30}
\end{eqnarray}
which naturally yields,
\begin{eqnarray}
x(u)=\int_{u}^{u_c}\frac{du}{\sqrt{\mathcal{H}(u)}}.
\label{E31}
\end{eqnarray}

Using (\ref{E30}), it is indeed quite trivial to show that,
\begin{eqnarray}
\frac{d \upsilon}{d u}=-\frac{1}{f(u)h(u)}\left( 1+\frac{\mathsf{E}h(u)}{\sqrt{\mathcal{H}(u)}}\left(1-\frac{f(u)\beta^{2}L^{4}}{u^{2}} \right)\right). 
\label{E32}
\end{eqnarray}

Next, combining (\ref{E17}) and (\ref{E31}) we note,
\begin{eqnarray}
\frac{\mathfrak{R}}{2}=\int_{u_{c}}^{u_T}\frac{u^{3}du}{\sqrt{c(u)}}\frac{1}{\sqrt{\mathsf{J}^{2}-\frac{u^{6}}{c(u)}}}+\int_{0}^{u_c}\frac{du}{\sqrt{\mathcal{H}(u)}}
\label{E33}
\end{eqnarray}
where, we have assumed that $ u(x) $ is a monotonically decreasing function of $ x $ \cite{Liu:2013qca}.

Moreover, integrating (\ref{E32}) we find,
\begin{eqnarray}
\mathfrak{t}=\int_{0}^{u_c}\frac{du}{f(u)h(u)}\left( 1+\frac{\mathsf{E}h(u)}{\sqrt{\mathcal{H}(u)}}\left(1-\frac{f(u)\beta^{2}L^{4}}{u^{2}} \right)\right). 
\label{E34}
\end{eqnarray}
Before we proceed further, the crucial point that is to be noted at this stage is the following: The integrand above in (\ref{E34}) clearly seems to be divergent near the horizon, $ u \sim u_H $ due to the vanishing of the function $ f(u) $ there. However, it turns out that,
\begin{eqnarray}
\lim_{u \rightarrow u_H} \left( 1+\frac{\mathsf{E}h(u)}{\sqrt{\mathcal{H}(u)}}\left(1-\frac{f(u)\beta^{2}L^{4}}{u^{2}} \right)\right)\approx 1+\frac{\mathsf{E}}{\sqrt{\mathcal{H}(u_H)}}.
\end{eqnarray}
Now, since $ \mathcal{H}(u_H)=\mathsf{E}^{2} $, therefore the above integral (\ref{E34}) is finite iff, $ \mathsf{E}<0 $, which in turn suggests that the second term on the R.H.S. of (\ref{E27}) must be less than the corresponding leading term. 

Our next natural task would be to find the area functional(s) corresponding to both $ \upsilon < 0 $ as well as $ \upsilon > 0 $ regions separately. The area functional (\ref{E10}) corresponding to  $ \upsilon <0 $ region turns out to be,
\begin{eqnarray}
\mathcal{A}_{-}=L_{y}L_{\chi}L^{3}\int_{\frac{u_c}{u_T}}^{1}  \frac{d\xi}{u_T^{2}\xi^{3}}\frac{\sqrt{c(\xi)}}{\sqrt{1-\frac{c_T}{c(\xi)}\xi^{6}}}
\end{eqnarray}
where, we have defined a new variable, $ \xi = \frac{u}{u_T}$ such that, $ u_T=(c_T \mathsf{J}^{2})^{1/6} $ corresponds to the so called terminal point along with, $ c_T=c(u_T) $.

Finally, the area functional corresponding to $ \upsilon >0 $ region turns out to be,
\begin{eqnarray}
\mathcal{A}_{+}=L_{y}L_{\chi}L^{3}\int_{0}^{\frac{u_c}{u_T}}\frac{d\xi}{\sqrt{c_T}u_T^{2}\xi^{6}}\frac{\sqrt{c(\xi)\mathcal{K}(\xi)}}{\sqrt{\mathcal{H}(\xi)}}
\end{eqnarray}
which eventually results in a total area functional of the form,
\begin{eqnarray}
\mathcal{A}=\frac{L_{y}L_{\chi}L^{3}}{u_T^{2}}\left[\int_{\frac{u_c}{u_T}}^{1}  \frac{d\xi}{\xi^{3}}\frac{\sqrt{c(\xi)}}{\sqrt{1-\frac{c_T}{c(\xi)}\xi^{6}}}+ \int_{0}^{\frac{u_c}{u_T}}\frac{d\xi}{\sqrt{c_T}\xi^{6}}\frac{\sqrt{c(\xi)\mathcal{K}(\xi)}}{\sqrt{\mathcal{H}(\xi)}}\right].
\label{E38}
\end{eqnarray}

Clearly, the integral (\ref{E38}) is divergent near the UV scale of the boundary theory and therefore it should be regularized by means of a proper UV cut off. On top of it, in our analysis, we would be mostly interested to explore the behaviour of the area functional (\ref{E38}) after a \textit{global} quantum quench when the system passes from the so called vacuum to the thermally excited state. As a result, the quantity that we would be mostly interested to compute is the change in the area functional namely, $ \Delta \mathcal{A}=\mathcal{A}-\mathcal{A}^{(vac)}_{-} $ where,
\begin{eqnarray}
\mathcal{A}^{(vac)}_{-} =L_{y}L_{\chi}L^{3}\int_{0}^{1}  \frac{d\xi}{u_T^{2}\xi^{3}}\frac{\sqrt{c(\xi)}}{\sqrt{1-\frac{c_T}{c(\xi)}\xi^{6}}}
\end{eqnarray}
is the extremal (hyper)surface corresponding to the vacuum configuration.


\subsection{Pre-local equilibrium}
We first focus on the behaviour of holographic entanglement entropy at early times namely, $  \mathfrak{t} \ll u_H $ such that, $ u_c/u_H \ll 1 $. Note that here $ u_H $ could be identified as the time scale for local equilibrium growth. However, during such pre local equilibrium stage, one could safely take the limit, $ u_c \rightarrow 0 $ where the extremal surface crosses the null shell. Under such circumstances, the space time region corresponding to the black brane configuration turns out to be extremely small.

To start with, from (\ref{E34}), we note that,
\begin{eqnarray}
\mathfrak{t}=\int_{0}^{\frac{u_c}{u_T}}\frac{u_T^{2}\xi d\xi}{\beta L^{2}}\left(1-\frac{\beta^{3}L^{6}\mathsf{E}}{\mathsf{J}}+\left( \frac{\mathsf{E}u_T^{2}\beta L^{2}}{\mathsf{J}}-\frac{u_T^{2}}{2\beta^{2}L^{4}} \left( 1-\frac{\beta^{4}L^{8}}{u_H^{4}}\right)\right) \xi^{2} \right) +\mathcal{O}(\xi^{5}).
\label{E39}
\end{eqnarray}

Performing the above integral (\ref{E39}), it is indeed quite trivial to find,
\begin{eqnarray}
\mathfrak{t}\approx \frac{u_c^{2}}{2\beta L^{2}}\left( 1-\frac{\beta^{3}L^{6}\mathsf{E}}{\mathsf{J}}-\frac{u_c^{2}}{4\beta^{2}L^{4}}\left(1-\frac{2 \mathsf{E}\beta^{3}L^{6}}{\mathsf{J}} \right) \right) +\frac{\beta L^{2}}{8}\left( \frac{u_c}{u_H}\right)^{4}
\label{E41}
\end{eqnarray}
which could be inverted to obtain,
\begin{eqnarray}
u_c \approx \frac{\sqrt{\mathfrak{t}}}{\sqrt{L}\mathfrak{a}}+\mathcal{O}(\mathfrak{t}^{3/2})
\label{E42}
\end{eqnarray}
where, $ \mathfrak{a}=\frac{1}{2\beta L^{2}} \left( 1-\frac{\beta^{3}L^{6}\mathsf{E}}{\mathsf{J}}\right) $.

Our next task would be to compute the difference, $ \Delta \mathcal{A}=\mathcal{A}-\mathcal{A}^{(vac)}_{-} $. Considering the limit $ u_c/u_T \ll 1 $, we finally obtain,
\begin{eqnarray}
|\Delta\mathcal{A}^{(Sch)}|\approx \frac{L_{y}L_{\chi}L c_T u_c^{5}}{10 u_T^{6}\beta}+\mathcal{O}((u_c/u_H)^{4}).
\label{E43}
\end{eqnarray}

Using (\ref{E42}) and (\ref{E43}), we note that,
\begin{eqnarray}
|\Delta\mathcal{A}^{(Sch)}| \sim \mathfrak{t}^{\frac{5}{2}}.
\label{E44}
\end{eqnarray}

Clearly, for non relativistic QFTs associated with Sch isometry group and corresponding to  $ z=2 $ fixed point we observe a non trivial time evolution of the area functional which is indeed different from their relativistic as well as Lifshitz cousins \cite{Liu:2013qca}-\cite{Fonda:2014ula}. A careful observation reveals that the above formula in (\ref{E44}) could be put in the form, $  \mathfrak{t}^{2+\frac{1}{z}}$ where the additional factor of time ($ \mathfrak{t} $) we identify as the contribution coming from the $ \chi $ integral (that precisely gives rise to an additional length factor $ L_{\chi} $) in the area functional (\ref{E38}).

\subsection{Post-local equilibrium}
The temperature scale that we focus in this section is $ \frac{\mathfrak{R}}{2}\gg \mathfrak{t}\gg u_H $. The typical gravity picture in this region is the following: In this regime of time, the hyper-surface starts penetrating the null shell at a scale ($ u_c^{\ast} $) that is bigger than the horizon radius itself namely, $ u_c^{\ast} \geq u_H $. Those extremal surfaces for which $ u_c <u_c^{\ast} $, reach the boundary. On the other hand, extremal surfaces for which $ u_c > u_c^{\ast} $ never reach the boundary and hit the singularity. Such extremal surfaces are called critical extremal surfaces and the corresponding point of intersection ($ u_c^{\ast} $) is known as the critical point \cite{Liu:2013qca}.

In order to study critical extremal surface, we first argue that Eq.(\ref{E30}) could be thought of as describing the motion of a particle in one dimension with an effective potential, $\mathsf{V}_{eff}(u)\equiv \mathcal{H}(u) $. The minima of this potential is guaranteed to produce a stable orbit which is obtained by tuning the parameter $ u_c $ such that both the velocity as well as the acceleration of the particle turns out to be zero at that point. This yields,
\begin{eqnarray}
\mathcal{H}'(u_m)|_{u_c=u^{\ast}_{c}}\sim 0 ,~~\mathcal{H}(u_m)|_{u_c=u^{\ast}_{c}}=0
\label{e49}
\end{eqnarray}
where, $ u=u_m(\ll u_T) $ is some point in between $ u_T $ and $ u_H $ that minimizes the potential \cite{Liu:2013qca}. 

In the following, we further illustrate the above discussions and provide a detailed calculation in order to locate this critical extremal surface. From (\ref{e28}), we first note that for an increment,  
\begin{eqnarray}
u_c \rightarrow \tilde{u}_c=2^{1/4}u_H(1+\varepsilon)\equiv u_s (1+\varepsilon),~~\varepsilon> 0,~~|\varepsilon |\ll 1
\end{eqnarray}
the change in $ u'_{+} $ turns out to be,
\begin{eqnarray}
\Delta u'_{+} \approx - 4\varepsilon u'_{-}\left(1+\frac{u'_{+}(u_s)}{2|u'_{-} |}\left(1+\frac{\mathcal{X}'(u_s)}{2\mathcal{X}(u_s)} \right)\right) +\mathcal{O}(\varepsilon^{2})\approx - 4\varepsilon u'_{-} >0 
\end{eqnarray}
where, $ \mathcal{X}(u_s) = \frac{1}{2\sqrt{g(u_s)}c(u_s)}\left( \frac{1}{\upsilon'^{(0)}}+\sqrt{g(u_s)}|u'_s|\right)$, and we have also considered the fact that, $u'_{+}(u_s) =0  $. Therefore, from the above discussion we note that the extremal surfaces those intersect the null shell at $ u_c > u_s $, always move away from the boundary and never contribute to the entanglement entropy of the boundary theory.

Next, from (\ref{E30}), we note that the first term in $ \mathcal{H}(u) $ is zero both at $ u=u_H $ and $ u=u_T/c_T^{1/6} $ and is negative for, $u_H <u<u_T/c_T^{1/6} $. Note that, in the large $ u_T (\gg u_H)$ limit, this upper bound saturates exactly at $ u_T $ \cite{Liu:2013qca}. As a consequence of this, $ \mathcal{H}(u) $ exhibits a minima in between which could be obtained from (\ref{e49}) as,
\begin{eqnarray}
u_T^{6}=\frac{c_T u_m^{7}f'(u_m)}{6 f(u_m)}\left( \frac{u_T^{6}}{c_T u_m^{6}}-1-\frac{\mathsf{E}^{2}\beta^{2}L^{4}}{u_m^{2}}\right) +\frac{c_T u_m^{4}\mathsf{E}^{2}\beta^{2}L^{4}}{3}.
\label{e52}
\end{eqnarray}

Substituting, $ u_m \sim u_s =2^{1/4}u_H $ into (\ref{e52}) we find,
\begin{eqnarray}
u_T^{(s)6}=\frac{2^{7/2}c^{(s)}_T u_H^{6}}{3}\left(\frac{u_T^{(s)6}}{2^{3/2}c^{(s)}_T u_H^{6}}-1-\frac{\mathsf{E}^{2}\beta^{2}L^{4}}{\sqrt{2}u_H^{2}} \right)+\frac{2c^{(s)}_T u_H^{4}\mathsf{E}^{2}\beta^{2}L^{4}}{3}.
\label{e53} 
\end{eqnarray}

On the other hand, from the second condition in (\ref{e49}) we obtain,
\begin{eqnarray}
\frac{u_T^{(s)6}}{2^{3/2}c^{(s)}_T u_H^{6}}=1+\mathsf{E}^{2}\left(1+\frac{\beta^{2}L^{4}}{\sqrt{2}u_H^{2}} \right).
\label{e54} 
\end{eqnarray}

Substituting (\ref{e54}) into (\ref{e53}), we find,
\begin{eqnarray}
u_T^{(s)}=\mathsf{E}^{1/3}\left(1+ \frac{2c^{(s)}_T u_H^{4}\beta^{2}L^{4}}{3}\right)^{1/6} 
\end{eqnarray}
which clearly indicates the existence of the tip corresponding to a family of extremal surfaces that includes the critical extremal surface parametrized by, $ u_c^{\ast}=u_m=u_s$.

Finally, in order to compute the area functional (\ref{E38}) in this region, we consider the expansion,
\begin{eqnarray}
u_c = u^{\ast}_{c} (1-\lambda) ,~~|\lambda |\ll 1.
\label{E47}
\end{eqnarray}
Under such circumstances, the major contribution to the area functional (\ref{E38}) comes from the sector, $ u \sim u_m \sim u_T \xi_{m} $. In this sector, one can expand,
\begin{eqnarray}
\mathcal{H}(\xi)&=&\mathcal{H}(\xi_{m})+(\xi -\xi_{m})\mathcal{H}'(\xi_{m})+\frac{1}{2}(\xi -\xi_{m})^{2}\mathcal{H}''(\xi_{m})+\mathcal{O}((\xi -\xi_{m})^{3})\nonumber\\
&\approx &(\xi -\xi_{m})\mathcal{H}'(\xi_{m})+\frac{1}{2}(\xi -\xi_{m})^{2}\mathcal{H}''(\xi_{m}).
\end{eqnarray} 

By virtue of (\ref{E47}), we note,
\begin{eqnarray}
(\xi -\xi_{m})\mathcal{H}'|_{\xi_{c}=\xi^{\ast}_{c}}\equiv (\xi_{c}-\xi^{\ast}_{c})\mathcal{H}'(\xi_{c})=-\xi^{\ast}_{c}\lambda \mathcal{H}'(\xi_{c})
\end{eqnarray}
such that, $ | \xi^{\ast}_{c}\lambda \mathcal{H}'(\xi_{c})|\ll 1$.

Under such circumstances, from (\ref{E34}) one arrives at,
\begin{eqnarray}
\mathfrak{t} &\approx & -\int_{\xi \sim \xi_{m}}\frac{u_T~\mathsf{E}^{\ast}~ d(\xi_{m}-\xi)}{f(\xi_{m})\sqrt{-\xi^{\ast}_{c}\lambda \mathcal{H}'(\xi_{c})+\frac{1}{2}(\xi_{m} -\xi)^{2}\mathcal{H}''(\xi_{m})}}\left(1-\frac{f(\xi_{m})\beta^{2}L^{4}}{\xi_{m}^{2}u_T^{2}} \right)  \nonumber\\
& \approx & -\frac{u_T~\mathsf{E}^{\ast}}{f(\xi_{m})\sqrt{\frac{1}{2}\mathcal{H}''(\xi_{m})}}\left(1-\frac{f(\xi_{m})\beta^{2}L^{4}}{\xi_{m}^{2}u_T^{2}} \right) ~ \log\lambda
\label{E50}
\end{eqnarray}
where,
\begin{eqnarray}
\mathsf{E}^{\ast}=-\frac{1}{\sqrt{c(u_m)\mathcal{K}(u_m)}}\sqrt{- f(u_m)\left( \frac{u_T^{6}}{c_Tu_m^{6}}-1\right) }.
\end{eqnarray}

From (\ref{E33}), the length of the strip turns out to be,
\begin{eqnarray}
\mathfrak{R}\approx \frac{u_T \sqrt{\mathfrak{b}}}{2} \text{Hypergeometric2F1}\left[\frac{1}{2},\frac{2}{3},\frac{5}{3},\mathfrak{b}\right]-\frac{2u_T}{\sqrt{\frac{1}{2}\mathcal{H}''(\xi_{m})}} ~ \log\lambda
\end{eqnarray}
where, we have set $ \mathfrak{b}=\frac{c_T}{c(\xi_{m})} $.

Finally, the area functional turns out to be,
\begin{eqnarray}
\Delta \mathcal{A}^{(Sch)}\approx -\frac{L_{y}L_{\chi}L^{3}\sqrt{\mathcal{K}(\xi_{m})}}{u_T^{2}\sqrt{\mathfrak{b}}\xi_{m}^{6}\sqrt{\frac{1}{2}\mathcal{H}''(\xi_{m})}}~\log\lambda.
\label{E53}
\end{eqnarray}

Combining, (\ref{E50}) and (\ref{E53}) we finally obtain,
\begin{eqnarray}
\Delta \mathcal{A}^{(Sch)} =\frac{L_{y}L_{\chi}L^{3}u_T^{3} f(u_{m})}{u_{m}^{6}\mathsf{E}^{\ast}\sqrt{c_T}\sqrt{c(u_{m})\mathcal{K}(u_{m})}}\mathfrak{t}.
\label{E54}
\end{eqnarray}

In the large $ u_T(\gg u_m) $ limit, one could further simplify (\ref{E54}) as,
\begin{eqnarray}
\Delta \mathcal{A}^{(Sch)} \approx \left( \frac{L_{y}L_{\chi}L^{3}}{u_H^{3}}\right) \mathfrak{v}^{(Sch)}_{s}\mathfrak{t}
\end{eqnarray}
where,
\begin{eqnarray}
\mathfrak{v}^{(Sch)}_{s} &=& \frac{\sqrt{- f(u_m)}}{\sqrt{c_T}}\left( \frac{u_H}{u_m}\right)^{3} \approx \mathfrak{v}^{(Rel)}_{s}\left(1 -\left( \frac{u_H}{u_T}\right) ^{4} \left(1-\frac{u_H^{4}}{\beta^{2}L^{4}u_T^{2}} \right) \right) \nonumber\\
\mathfrak{v}^{(Rel)}_{s} &=& \sqrt{- f(u_m)}\left( \frac{u_H}{u_m}\right)^{3} .
\label{E56}
\end{eqnarray}
Here, $ \mathfrak{v}^{(Sch)}_{s} $ is the so called tsunami velocity associated with Schr{\"o}dinger $ Dp $ branes in the bulk and corresponds to the \textit{linear} growth of the entanglement entropy (during the post local equilibrium stage) which is independent of the shape of the entangling region \cite{Liu:2013qca}. From (\ref{E56}), it is also evident that in the large $ u_T (\gg u_H)$ limit, the tsunami velocity in a non relativistic set up  gradually saturates to its corresponding value in the relativistic set up ($ \mathfrak{v}^{(Rel)}_{s} $) \cite{Liu:2013qca}. 

\subsection{Saturation}
The situation that we consider in this section essentially corresponds to the fact that the extremal surface lies entirely outside the horizon of the black brane namely, $ u_T\leq u_H $. In this region, the entanglement entropy is supposed get saturated to its equilibrium value and which essentially turns out to be the entropy corresponding to a thermal state. As a result, in this regime, the major contribution to the entanglement entropy arises from the geometry around the horizon of the black brane.

 In order to proceed further we set, 
\begin{eqnarray}
u_c = u_T (1-\lambda),~~u_T \sim u_H.
\label{E58}
\end{eqnarray}

Using (\ref{E58}), it is trivial to show,
\begin{eqnarray}
\frac{\mathfrak{R}}{2}\approx \int_{0}^{1-\lambda}\frac{u_H d\xi}{\sqrt{\mathcal{H}(\xi)}}.
\label{E59}
\end{eqnarray}

On the other hand, from (\ref{E38}) we note that,
\begin{eqnarray}
\mathcal{A}\approx \frac{L_{y}L_{\chi}L^{3}}{\sqrt{c_T}u_H^{2}} \int_{0}^{1-\lambda}\frac{d\xi}{\xi^{6}}\frac{\sqrt{c(\xi)\mathcal{K}(\xi)}}{\sqrt{\mathcal{H}(\xi)}}.
\label{E60}
\end{eqnarray}

Using (\ref{E59}), one could further rewrite (\ref{E60}) as,
\begin{eqnarray}
\mathcal{A}^{(Sch)}_{sat}&\approx & \frac{L_{y}L_{\chi}L^{3}\mathfrak{R}}{2\sqrt{c_T}u_H^{3}}+\frac{L_{y}L_{\chi}L^{3}}{\sqrt{c_T}u_H^{2}} \int_{\frac{\epsilon}{u_H}}^{1}\frac{d\xi}{\xi^{6}}\frac{\sqrt{c(\xi)\mathcal{K}(\xi)}}{\sqrt{\mathcal{H}(\xi)}}\left(1-\frac{\xi^{6}}{\sqrt{c(\xi)\mathcal{K}(\xi)}} \right)\nonumber\\
&\approx &\frac{L_{y}L_{\chi}L^{3}\mathfrak{R}}{2\sqrt{c_T}u_H^{3}}+\frac{\beta L_{y}L_{\chi}L^{5}}{3 \epsilon^{3}}+..~..
\label{E61}
\end{eqnarray}
where, $ \epsilon $ is the so called UV cut-off of the theory.    

We now focus towards estimating the saturation time ($ \mathfrak{t}_{sat} $) for the entanglement entropy in the theory. To do that, we first consider the case of \textit{continuous} transitions where the entanglement entropy is continuous across the saturation time, $ \mathfrak{t}=\mathfrak{t}_{sat} $. 

 From (\ref{E29}), we first note that,
 \begin{eqnarray}
 \upsilon'^{(0)}|_{u_c \sim u_H} \sim \frac{1}{f(u_H)} \rightarrow \infty
 \end{eqnarray}
which together with the fact, $ u_c' \sim u_T' \sim 0 $ naturally yields, 
\begin{eqnarray}
\mathsf{E}|_{u_c \sim u_H} \approx 0.
\end{eqnarray}

As a next step of our analysis, we expand the function $ \mathcal{H}(\xi) $ about $ \xi \sim \xi_{H} $,
\begin{eqnarray}
\mathcal{H}(\xi) \approx -\lambda \xi_{H}\mathcal{H}'(\xi_{H})+\frac{1}{2}(\xi -\xi_{H})^{2}\mathcal{H}''(\xi_{H}).
\label{E64}
\end{eqnarray}

Substituting (\ref{E64}) into (\ref{E59}), we finally obtain,
\begin{eqnarray}
\frac{\mathfrak{R}}{2}\approx -u_H \int_{\xi \sim \xi_{c}}\frac{d(\xi_{H}-\xi)}{(\xi_{H}-\xi)}\sqrt{\frac{2}{\mathcal{H}''(\xi_{H})}}\approx -\frac{\sqrt{2}u^{2}_H}{\beta L^{2}\sqrt{|f''(\xi_{H})}|}\log \lambda.
\label{E65}
\end{eqnarray}

Finally, the saturation time turns out to be,
\begin{eqnarray}
\mathfrak{t}_{sat}\approx - \int_{\xi \sim \xi_{c}}\frac{u_H d(\xi_{H}-\xi)}{|f'(\xi_{H})|(\xi_{H}-\xi )}=-\frac{u_H }{|f'(\xi_{H})|}\log \lambda.
\label{E66}
\end{eqnarray}

Comparing (\ref{E65}) and (\ref{E66}), we finally arrive at,
\begin{eqnarray}
\mathfrak{t}_{sat}\approx \left( \frac{\sqrt{6}\beta L^{2}}{8u_H}\right) \mathfrak{R}.
\label{E67}
\end{eqnarray}

In case of \textit{discontinuous} transition, there is no straightforward formula for the saturation time. However, considering the fact that the area functional is still continuous across the transition point, one might try to have a rough estimate on the characteristic time scale for saturation. Considering linear growth all the way upto saturation point we find,
\begin{eqnarray}
\mathfrak{t}_{l}\approx \frac{\mathfrak{R}}{2 \sqrt{c_T}\mathfrak{v}^{(Sch)}_{s}}+\mathcal{O}(u_H/\mathfrak{R}).
\label{E68}
\end{eqnarray}

Comparing (\ref{E67}) and (\ref{E68}) we find,
\begin{eqnarray}
\frac{\mathfrak{t}_{l}}{\mathfrak{t}_{sat}}\approx \frac{4u_H}{\sqrt{6}\beta L^{2}}\left(\frac{u_m}{u_H} \right) >1
\end{eqnarray}
which thereby suggests that in the case of discontinuous saturation, the linear growth might persists for a longer time compared to that of the case of a continuous saturation \cite{Liu:2013qca}.

As a final goal of our analysis, we would like to explore the conditions for a saturation to be continuous in the context of non relativistic quench. For the saturation to be continuous, the entity, $ \mathfrak{t}-\mathfrak{t}_{sat} $ must be negative all the way to, $ u_c-u_T \rightarrow 0 $ \cite{Liu:2013qca}.

To start with we note that,
\begin{eqnarray}
u_c = u_T (1-\lambda),~~u_T = u_B (1-\varsigma),~~|\lambda| , |\varsigma| \ll 1
\end{eqnarray}
where we consider large entangling region such that, $ u_B \sim u_H $.

Under such circumstances, from (\ref{E27}) we note that,
\begin{eqnarray}
\mathsf{E}\approx -\frac{\sqrt{6\lambda g(u_T)}}{2}.
\end{eqnarray}

Finally, the difference turns out to be,
\begin{eqnarray}
\mathfrak{t}-\mathfrak{t}_{sat}\approx - \frac{\mathfrak{K}u_T\sqrt{6g(u_T)}}{2}\sqrt{\lambda}+\mathcal{O}(\lambda)
\end{eqnarray}
where,
\begin{eqnarray}
\mathfrak{K} =\int_{0}^{1}\frac{d\xi}{f(\xi)}\left(1-\frac{f(\xi)\beta^{2}L^{4}}{\xi^{2}u_T^{2}} \right)\frac{\sqrt{c_T}\xi^{3}}{\sqrt{f(\xi)(1-c_T \xi^{6})}}.
\label{E75}
\end{eqnarray}
Clearly, the saturation is continuous for $ \mathfrak{K}>0 $ and on the other hand it is discontinuous for $ \mathfrak{K}<0 $. Computing the integral above in (\ref{E75}), we find,
\begin{eqnarray}
\mathfrak{K} \approx \frac{\beta L^{2}}{2u_H}\left(1-\frac{2\beta^{2}L^{4}}{u_H^{2}} \left( 1-\frac{\beta^{2}L^{4}}{6u_H^{2}}\right) \right). 
\end{eqnarray}
If we now set the dimensionless quantity, $ \frac{\beta^{2}L^{4}}{u_H^{2}}\equiv \mathfrak{n} $, where $ \mathfrak{n} $ is some real number ($ \mathfrak{n}\in \mathbb{R} $), then the condition for the saturation to be discontinuous turns out to be,
\begin{eqnarray}
\frac{\beta^{2}L^{4}}{u_H^{2}} \equiv \mathfrak{n}\leq 3 +\sqrt{6}
\end{eqnarray}
which thereby imposes an important constraint on the parameters of the theory.

\section{Summary and final remarks}
We now summarize the key findings of our analysis. In this paper, based on the holographic techniques, we explore the physics of thermalization for special class of strongly interacting QFTs those are dual to Schr{\"o}dinger $ Dp $ brane configurations in the bulk. Such QFTs typically correspond to system of fermions at unitarity \cite{Son:2008ye}. Therefore what this paper explores is the fate of thermalization corresponding to a system of fermions at unitarity (popularly known as cold atom systems).

In order to probe such \textit{non equilibrium} processes, we explore the time evolution of (holographic) entanglement entropy during the \textit{global} thermal quench in a strongly coupled medium. We perform our analysis at three different time scales. It turns out that during the very early stages of thermal evolution, namely the pre local equilibrium growth, the entanglement entropy exhibits a faster growth in time ($ \Delta \mathcal{S}_{EE}\sim \mathfrak{t}^{5/2} $) as compared to those of its relativistic as well as Lifshitz cousins \cite{Liu:2013qca}-\cite{Fonda:2014ula}. In fact, this has been the fastest growth of entanglement entropy as observed so far. However, during the post local equilibrium growth, the most dominant contribution to the area functional seems to be appearing form so called critical extremal surface which yields a linear growth for the entanglement entropy. Therefore, as far as the post local growth is concerned, non relativistic QFTs with Schr{\"o}dinger isometry group are no different from their relativistic cousins.  On top of it, as an additional observation, we note that the so called tsunami velocity always saturates the corresponding relativistic bound \cite{Liu:2013qca}. This is quite surprising from the point of view of a non relativistic theory as in principle there should not be any upper bound on the speed of propagation of the tsunami wave. Finally, at times much lager than the time scale corresponding to thermal saturations, the entanglement entropy reaches its saturation point and the most dominant contribution to the entanglement entropy seems to be appearing from the extremal surface around the horizon of the black brane. We also probe in deep conditions for such saturation to occur. 

{\bf {Acknowledgements :}}
 The author would like to acknowledge the financial support from UGC (Project No UGC/PHY/2014236).


\end{document}